\documentclass[prx, twocolumn, superscriptaddress, dvipsnames]{revtex4-2}
\pdfoutput=1
\usepackage[utf8]{inputenc}
\usepackage[english]{babel}
\usepackage{amsmath}
\usepackage{amssymb}
\usepackage{amsthm}
\usepackage{amsfonts}
\usepackage{bbm}
\usepackage{makecell}
\usepackage{xcolor}
\usepackage{bm}
\usepackage{here}
\usepackage{soul}

\usepackage{braket}
\usepackage{bbold}
\usepackage[utf8]{inputenc}
\usepackage{qcircuit}

\usepackage{xcolor}
\definecolor{orangered}{RGB}{255, 69, 0}
\xyoption{color}

\usepackage[colorlinks,allcolors=blue]{hyperref}
\usepackage[noabbrev, capitalize, nameinlink]{cleveref}

\creflabelformat{equation}{#2\textup{#1}#3}
\theoremstyle{definition}

\usepackage{tikz}
\usetikzlibrary{decorations.pathreplacing,calligraphy,decorations.markings}

\definecolor{tensorcolor}{rgb}{0.65,0.77,0.95}
\definecolor{mpdocolor}{HTML}{FCDE70}
\definecolor{whampdocolor}{HTML}{FCDE70}
\definecolor{whampdocolorw}{HTML}{EEF7FF}
\definecolor{mpdotcolor}{rgb}{1,0.98,0.94}
\definecolor{btensorcolor}{rgb}{0.65,0.50,0.69}
\definecolor{whitetensorcolor}{HTML}{F8F8F8}
\definecolor{diamondcolor}{HTML}{E6F1ED}
\definecolor{unitarycolor}{rgb}{0.8,0.5,.5}
\definecolor{lcolor}{HTML}{D9EAFD}

\newcommand\mthick{}

\newcommand{\ancilla}[5]{
\begin{scope}[shift={(#1)}]
\def\sy{0.5};
\pgfmathsetmacro{\offset}{-0.5*(#2-1)-1};
\foreach \x in {1,..., #2} {
\pgfmathsetmacro{\xo}{(\offset+\x)*#4};
        \draw[\mthick] (\xo,-#3) -- (\xo,#3);
        \draw (\xo,-#3-\sy) node {\small $|0\rangle$};
        \ifnum#5=1
        \draw (\xo,#3+\sy) node {\small $\langle 0|$};
        \fi
}
\end{scope}
}

\newcommand{\hancilla}[5]{
\begin{scope}[shift={(#1)}]
\def\sy{0.5};
\pgfmathsetmacro{\offset}{-0.5*(#2-1)-1};
\foreach \x in {1,..., #2} {
\pgfmathsetmacro{\xo}{(\offset+\x)*#4};
        \draw[\mthick] (-#3,\xo) -- (#3,\xo);
        \draw (-#3-\sy,\xo) node {\small $|0\rangle$};
        \ifnum#5=1
        \draw (#3+\sy, \xo) node {\small $\langle 0|$};
        \fi
}
\end{scope}
}

\usetikzlibrary{decorations.pathmorphing, decorations.markings, arrows, arrows.meta, shapes, patterns}
\tikzset{baseline={([yshift=-.5ex]current bounding box.center)}}
\tikzset{every path/.style={ line width=0.5pt, line cap=round }}
\colorlet{Virtual}{RedOrange}
\tikzstyle{bevel} = [ preaction = { draw, white, line width=3pt,  line cap = round } ]
\tikzstyle{bevel wide} = [ preaction = { draw, white, line width=4pt,  line cap = round } ]
\tikzstyle{symb} = [ draw=black, fill=black, line width=0.4pt, inner sep=1.5pt ]
\tikzstyle{mysymb} = [ draw=black, fill=white, circle, line width=0.3pt, inner sep=1pt, font=\small ] 
\tikzstyle{symb large} = [ inner sep=2.1pt ]
\tikzstyle{symb small} = [ inner sep=1pt   ]
\tikzstyle{symb tiny} = [ inner sep=0.8pt ]
\tikzstyle{symb fdisk} = [ circle ]
\tikzstyle{symb disk} = [ circle ]
\tikzstyle{symb square} = [ rectangle ]
\tikzstyle{symb fsquare} = [ rectangle ]
\tikzstyle{Msymb}=[draw=black, fill=whampdocolor, circle, inner sep=1pt, font=\small]
\tikzstyle{Nsymb}=[draw=black, fill=whampdocolorw, circle, inner sep=1pt, font=\small]
\tikzstyle{-mid} = [ decoration={ markings, mark = at position 0.50*\pgfdecoratedpathlength+0.6*3pt with \arrow{>[width=2pt]} }, postaction={decorate} ]
\tikzstyle{mid-} = [ decoration={ markings, mark = at position 0.50*\pgfdecoratedpathlength+0.6*3pt with \arrow{<[width=2pt]} }, postaction={decorate} ]

\newcommand\subsetsim{\mathrel{%
  \ooalign{\raise0.2ex\hbox{$\subset$}\cr\hidewidth\raise-0.8ex\hbox{\scalebox{0.9}{$\sim$}}\hidewidth\cr}}}

\newcommand{\tr}{\mathrm{Tr}}
\newcommand{\bo}{\mathbbm{1}}

\newcommand{\dg}{\dagger}
\newcommand{\drangle}{\rangle\!\rangle}
\newcommand{\dlangle}{\langle\!\langle}

\definecolor{yuhan}{rgb}{0.9, 0, 0.5}

\allowdisplaybreaks

\begin{document}
\title{Polynomial-time certification of fidelity for many-body mixed states and mixed-state universality classes}

\author{Yuhan Liu}
\affiliation{Max Planck Institute of Quantum Optics, Hans-Kopfermann-Str. 1, Garching 85748, Germany}
\affiliation{Munich Center for Quantum Science and Technology (MCQST), Schellingstr. 4, 80799 M{\"{u}}nchen, Germany}

\author{Yijian Zou}
\affiliation{Perimeter Institute for Theoretical Physics, Waterloo, Ontario, Canada N2L2Y5}

\date{\today}

\begin{abstract}
Computation of Uhlmann fidelity between many-body mixed states generally involves full diagonalization of exponentially large matrices. In this work, we introduce a polynomial-time algorithm to compute certified lower and upper bounds for the fidelity between matrix product density operators (MPDOs). Our method maps the fidelity estimation problem to a variational optimization of sequential quantum circuits, allowing for systematic improvement of the lower bounds by increasing the circuit depth. Complementarily, we obtain certified upper bounds on fidelity by variational lower bounds on the trace distance through the same framework. We demonstrate the power of this approach with two examples: fidelity correlators in critical mixed states, and codeword distinguishability in an approximate quantum error-correcting code. Remarkably, the variational lower bound accurately track the universal scaling behavior of the fidelity with a size-consistent relative error, allowing for the extraction of previously unknown critical exponents. Our results offer an exponential improvement in precision over known moment-based bounds and establish a scalable framework for the verification of many-body quantum systems.
\end{abstract}
\maketitle

\textit{Introduction--}
The rapid development of quantum simulation offers an unprecedented ability to engineer and probe complex quantum many-body systems. A crucial task is verification: evaluating the quality of the prepared state against a theoretical target. The quantitative measure of success is the fidelity function, $F(\rho, \sigma) := \| \sqrt{\rho}\sqrt{\sigma} \|_1$, which captures the information-theoretic distance between the desired state $\sigma$ and the experimentally prepared state $\rho$. Standard approaches, including direct fidelity estimation~\cite{flammia2011direct}, variational fidelity estimation~\cite{Cerezo2020variational,Tan2021variational,Chen2021variational}, and classical shadow tomography~\cite{Huang_2020}, typically require at least one state to be pure or low-rank~\cite{Wang_2023}. This restriction precludes the verification of thermal states or dissipative many-body states, which are inherently high-rank.

Fidelity estimation also plays a pivotal role in understanding mixed-state phases of matter. Recently, extensive studies have shown that the interplay between long-range quantum entanglement and decoherence leads to a wide range of mixed-state phases beyond traditional pure-state classification \cite{coser2019classification,sang2024mixed,zou2023channeling,zhang2025probingmixedstatephasesquantum,sang2024approximate,markov,ma2023average,ma2023topological,sohal2024noisy,ellison2025toward,yang2025topologicalmixedstatesphases,zhang2022strange,lessa2025strong,lee2024exact,Lee2025symmetryprotected,wang2023intrinsic,de2022symmetry,vijay2025informationcriticalphasesdecoherence,sang2025mixedstatephaseslocalreversibility,fan2024diagnostics,Guo2025PRX,Lu_2023,Sun_2025,sala2024spontaneousstrongsymmetrybreaking}. The key characterizations of these phases involve information-theoretic quantities such as the fidelity function, trace distance, and conditional mutual information. However, numerical algorithms for efficiently computing these quantities are severely limited. Existing approaches are largely restricted to fine-tuned models admitting mappings to statistical mechanics, enabling semi-analytical solutions~\cite{fan2024diagnostics,markov}. The only known general numerical method involves computing the square root or logarithm of the density matrix through full diagonalization, requiring exponential computational resources and thus limiting the exploration of mixed-state phases beyond the fined-tuned models. Crucially, a fundamental feature of many of these states remains unexploited: they can be naturally represented by tensor networks—such as Matrix Product Density Operators (MPDOs) \cite{Verstraete_2004,Zwolak_2004,cirac2017matrix,kato2024exact,ruiz2024matrix,liu2025parent,liu2025trading} and their higher-dimensional generalizations—as they typically arise from finite-depth channels acting on a matrix product state (MPS) or a projected entangled pair state.

In this work, we bridge this gap by developing a polynomial-time algorithm to compute the fidelity between two mixed states represented as MPDOs. We map the fidelity estimation problem to a variational optimization of sequential quantum circuits via Uhlmann's theorem \cite{uhlmann1976transition,Hauru_2018}. Our method provides certified lower bounds on fidelity that can be systematically improved by increasing the circuit depth. We show that, in analytically tractable limiting cases, the optimal circuit is achievable with finite depth. Furthermore, we complement this with certified upper bounds derived from variational estimates of the trace distance. We demonstrate the power of this method on two physically motivated examples where accurate fidelity estimation was crucial but previously intractable. First, we analyze fidelity correlators, a recently proposed order parameter for strong-to-weak spontaneous symmetry breaking (SWSSB) \cite{lessa2025strong}, for critical mixed states. Our lower bounds certify that dephasing noise induces a notably slower decay of these correlators for system sizes up to $N=32$ spins, with distinct critical exponents from the noiseless correlator, thus indicating a different universality class for the decohered state. Second, we evaluate the distinguishability of low-energy eigenstates in critical systems under extensive decoherence, providing a diagnostic for the error decodability of the approximate QEC code \cite{yi2024complexity,sang2024approximate}. We find that our lower bounds accurately capture the system-size scaling of the fidelity, distinguishing between the correctable and uncorrectable noise. Remarkably, the variational bounds accurately track the universal scaling behavior with a size-consistent relative error, enabling extraction of critical exponents in mixed states. In both cases, our algorithm offers an exponential improvement in precision over known moment-based bounds in terms of subfidelity and superfidelity.

\textit{Lower bound by variational circuit} -- We consider two density operators $\rho,\sigma$ on the many-body Hilbert space $\mathcal{H}_s$. One of the most prominent measures of similarity between these two states is the \textit{fidelity function}, defined as
\begin{equation}
\label{eq:F_def}
    F(\rho,\sigma)=\| \sqrt{\rho}\sqrt{\sigma} \|_1.
\end{equation}
Computing fidelity is, in general, difficult, requiring diagonalization of all eigenvalues of the density matrices. The celebrated \textit{Uhlmann's theorem}~\cite{uhlmann1976transition} connects Eq.~\eqref{eq:F_def} to the overlap between purifications of the density matrices. Let $|\psi_\rho\rangle\!\rangle\in \mathcal{H}_s\otimes\mathcal{H}_p$ be a purification of $\rho$, i.e., $\tr_{\mathcal{H}_p}|\psi_\rho\drangle\dlangle\psi_\rho|=\rho$, where $\mathcal{H}_p$ is the purification Hilbert space whose dimension is not smaller than the rank of $\sigma$. The theorem states that
\begin{equation}
\begin{aligned}
    F(\rho,\sigma)&=\max\lbrace |\dlangle \psi_\rho|\psi_\sigma\drangle|:\\
    &\quad|\psi_\sigma\drangle\in \mathcal{H}_s\otimes\mathcal{H}_p,\tr_{\mathcal{H}_p}|\psi_\sigma\drangle\dlangle\psi_\sigma|=\sigma  \rbrace.
\end{aligned}
\end{equation}
Importantly, different purifications are related through unitary equivalence. Let $|\psi_\sigma\drangle,|\psi'_{\sigma}\drangle\in \mathcal{H}_s\otimes \mathcal{H}_p$ be two purifications of $\sigma$, then there exists a unitary operator $U\in\mathrm{U}(\mathcal{H}_p)$ such that $|\psi'_{\sigma}\drangle=(\bo\otimes U)|\psi_\sigma\drangle$. Therefore, given a  purification $|\psi_\rho\drangle$ of $\rho$ and $|\psi_\sigma\drangle$ of $\sigma$, the fidelity function can be evaluated as an optimization problem
\begin{equation}
\label{eq:main_opti}
\begin{aligned}
    F(\rho,\sigma)&=\max_{U\in\mathrm{U}(\mathcal{H}_p)} |\dlangle \psi_\rho|(\bo\otimes U)|\psi_\sigma\drangle|.
\end{aligned}
\end{equation}
Any unitary $U$ thus provides a rigorous lower bound of the fidelity. 

\begin{figure*}
    \includegraphics[width=0.96\textwidth]{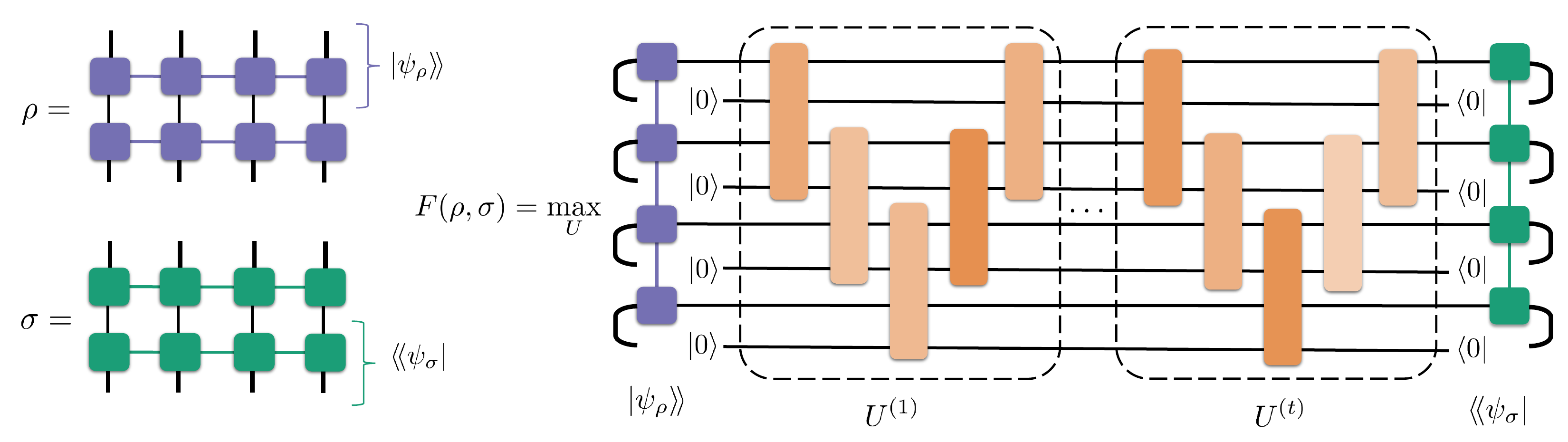}
    \caption{Fidelity of matrix product density operators from sequential circuit optimization. Left: $\rho = \tr_{\mathcal{H}_p} |\psi_{\rho}\drangle\dlangle \psi_{\rho}|$ is a locally purified MPDO. The upper part of the diagram represents the purification $|\psi_\rho\rangle\!\rangle$ as a two-sided MPS, where the bold vertical lines denote degrees of freedom in $\mathcal{H}_s$, and the thin vertical lines denote degrees of freedom in $\mathcal{H}_p$. Similar for $\sigma=\tr_{\mathcal{H}_p} |\psi_{\sigma}\drangle\dlangle \psi_{\sigma}|$ in the lower part. Right: Variational optimization of fidelity $F(\rho,\sigma)$ with a depth-$t$ sequential circuit with ancilla, with $U=U^{(t)}\cdots U^{(2)} U^{(1)}$. }
    \label{fig:ansatz}
\end{figure*}

To proceed, we require that $\rho$ and $\sigma$ are MPDOs. We further require that they acquire a local purification, i.e., $|\psi_\rho\drangle$ and $|\psi_\sigma\drangle$ are given as two-sided MPS, see Fig.~\ref{fig:ansatz}. This representation ensures the density matrices to be positive-semidefinite, a condition that is otherwise hard to verify~\cite{kliesch2014matrix}. Locally purified MPDOs also naturally appear in both equilibrium and non-equilibrium settings, such as thermal states of local Hamiltonians \cite{Zwolak_2004,Verstraete_2004,molnar2015approximating,Nguyen_2018,kato2019quantum,chen2020matrix}, boundary states of gappable 2D topological order \cite{molnar2022matrix}, and MPS under local Lindbladian evolutions \cite{Cheng_2021,Guo_2024,Godinez2025}. Instead of optimizing over all possible unitary $U$ in Eq.~\eqref{eq:main_opti}, we will use a variational circuit ansatz which allows efficient computation, and demonstrate that it also gives precise estimations of the fidelity in several examples.

\textit{Sequential circuit ansatz--} 
We adopt a sequential circuit ansatz \cite{Chen_2024}, as shown in Fig.~\ref{fig:ansatz}. We first add an extra ancilla degree of freedom initialized to $|0\rangle$ on each site of $\mathcal{H}_p$ for both purified states. Adding these ancilla still keeps the LPDOs as purification of $\rho$ and $\sigma$. We then apply a depth-$t$ sequential circuit $U=U^{(t)}\cdots U^{(2)} U^{(1)}$, where each local unitary in $U^{(i)}$ acts on two neighboring purified and ancillary degrees of freedom. In this paper, we adopt the convention that a depth-1 sequential circuit $U^{(i)}$ includes one forward layer of and one backward layer of untaries, see the dashed box of the right of Fig.~\ref{fig:ansatz}. A lower bound of the fidelity is then obtained by optimization over the variational circuit $U$ \footnote{In practice, the optimization is performed with quimb \cite{Gray2018} and pytorch through automatic differentiation, with optimal contraction order computed through cotengra \cite{Gray2021hyperoptimized}.}. Furthermore, systematic improvement can be achieved by increasing $t$.

To motivate our ansatz, we note that the action of the variational circuit can be implemented efficiently on the MPDO when the $t$ is finite. This ensures the polynomial-time complexity of our algorithm. Moreover, our ansatz captures known classes of matrix product unitary (MPU) going beyond FDLUs~\cite{schuch2011classifying,chen2011classification}.  Firstly, our ansatz captures all MPUs that are uniform (translation invariant). As shown in Ref.~\cite{cirac2017matrix,piroli2020quantum}, such MPUs are always quantum cellular automata (QCA) that preserve locality. They can be decomposed into FDLU and finite translation, both of which can be represented as finite-$t$ sequential circuits. Secondly, the addition of the ancilla further allows us to realize non-uniform MPUs that go beyond QCA. One example is the multi-control unitary (MCU), where a unitary is applied to the last qubit if the first $N-1$ qubits are in $|1\rangle$. As shown in SM, MCU can be achieved with $t=1$ sequential unitary with ancilla, consistent with linear-depth decomposition discussed in Ref.~\cite{zindorf2025efficient}.

\textit{Finite-$t$ optimal unitary.--} Our sequential circuit ansatz captures the optimal unitary in Eq.~\eqref{eq:main_opti} for several tractable examples. First, let $\rho = |\psi\rangle\langle\psi|$ with purification given by $|\psi_{\rho}\drangle = |\psi\rangle_s \otimes|0^{\otimes N}\rangle_p$. The fidelity is simply $F(\rho,\sigma) = \sqrt{\langle \psi |\sigma|\psi\rangle}$. For illustration, consider $\sigma = 2^{-N} \bo^{\otimes N}$, with the purification $|\psi_{\sigma}\drangle = \left( \frac{1}{\sqrt{2}}(|0_s0_p\rangle + |1_s 1_p\rangle) \right)^{\otimes N}$. The fidelity is $F(\rho,\sigma) = 2^{-N/2}$ and Uhlmann's theorem gives $F(\rho,\sigma) = 2^{-N/2}\max_{U} |\langle \psi|U|0^{\otimes N}\rangle|$. The maximum can be achieved with our sequential circuit ansatz, since every MPS of finite bond dimension can be prepared by a finite-depth sequential circuit from the product state \cite{schon2005sequential}.  More generally, when $\sigma$ is an MPDO with local purification, the optimial fidelity can still be obtained with our sequential circuit ansatz, see SM for details. 

Second, we consider examples where a finite-depth sequential circuit can achieve optimal fidelity while FDLUs cannot. Consider two mixed states $\rho,\sigma$ in the logical subspace $\mathcal{H}_L$ of the repetition code spanned by $|\bar{0}\rangle :=|0^{\otimes N}\rangle$ and $|\bar{1}\rangle :=|1^{\otimes N}\rangle$. Let $\rho = \sum_{i,j} \bar{\rho}_{ij} |\bar{i}\rangle \langle \bar{j}|$ and $\sigma = \sum_{i,j} \bar{\sigma}_{ij} |\bar{i}\rangle \langle \bar{j}|$, then the fidelity is $F(\rho,\sigma) = F(\bar{\rho},\bar{\sigma})$. Given any purification $|\psi_{\rho}\drangle, |\psi_{\sigma}\drangle \in \mathcal{H}_L \otimes \mathcal{H}_L $, the optimal unitary $U$ must perform a logical rotation, i.e., $U|\bar{j}\rangle = \sum_{i} \bar{U}_{ij} |\bar{i}\rangle $, where $\bar{U}\in SU(2)$. The logical unitary can be achieved with a finite-$t$ sequential unitary (see SM), but it cannot be realized as an FDLU due to Lieb-Robinson bounds.

\textit{Upper bound by trace distance} -- An upper bound of fidelity $F(\rho,\sigma)$ can be obtained using the Fuchs-van de Graaf inequality,
\begin{equation}
\label{eq:vandegraaf}
    F(\rho,\sigma)\leq \sqrt{1-\frac{1}{4}\|\rho-\sigma\|_1^2},
\end{equation}
where a certified lower bound of $\|\rho-\sigma\|_1$ can be obtained similarly using a variational unitary,
\begin{equation}
    \|\rho-\sigma\|_1=\max_{U\in \mathrm{U}(\mathcal{H}_s)} |\tr[U^\dg(\rho-\sigma)]|.
\end{equation}
By restricting $U$ to be the depth-$t$ sequential circuit ansatz, we then obtain a certified upper bound for the fidelity $F(\rho,\sigma)$.

Let us comment on the comparison to known bounds \cite{miszczak2008subsuperfidelityboundsquantum}
\begin{equation}
    E(\rho,\sigma)\leq F^2(\rho,\sigma)\leq G(\rho,\sigma)
\end{equation}
for generic mixed states, where $E(\rho_1, \rho_2) = \tr (\rho_1\rho_2) + \sqrt{2} \sqrt{(\tr(\rho_1\rho_2))^2 - \tr(\rho_1\rho_2 \rho_1 \rho_2)} $ is the subfidelity and $G(\rho_1,\rho_2) = \tr(\rho_1\rho_2) + \sqrt{(1-\tr\rho^2_1)(1-\tr \rho^2_2)}$ is the superfidelity. The quantities are efficiently computable for MPDOs as they only involve moments. However, the bound becomes exponentially loose, i.e., $E(\rho,\sigma) = e^{-\Theta(N)}$ and $G(\rho,\sigma) = 1-e^{-\Theta(N)}$, if the states have volume-law entropy. In the following, we demonstrate the exponential advantage of our approach for the applications to mixed-state phases and quantum error correction, where volume-law entropy naturally appears because of decoherence.

\textit{Fidelity correlators in the decohered critical state --} Strong-to-weak spontaneous symmetry breaking (SWSSB) is a recently proposed pattern of symmetry breaking for mixed states. For simplicity, we focus on the $\mathbb{Z}_2$ symmetry generated by the on-site symmetry operator $g =\prod_{i=1}^N Z_i$ of spin chains. The SWSSB order is characterized by the fidelity correlator $D_{ij}:=F(\rho, X_j X_i\rho X_i X_j)$.  Our variational framework provides a rigorous numerical route to compute bounds on the fidelity correlator, enabling the exploration of mixed-state universality classes in regimes inaccessible to exact diagonalization.

As a concrete example, 
we consider the low-energy eigenstates $|\psi^{(m)}\rangle$ of the Ising model under periodic boundary condition $H = - \sum_{i=1}^N ( X_i X_{i+1} + hZ_i)$ and apply the uniform $Z$-dephasing noise with strength $q$, $\mathcal{N}_i(\rho)=(1-\frac{q}{2})\rho+\frac{q}{2}Z_i\rho Z_i$ to obtain the mixed state $\rho^{(m)}=(\prod_i \mathcal{N}_i)|\psi^{(m)}\rangle\langle\psi^{(m)}|$ \cite{zou2023channeling,Lee2023quantum}. We compute the fidelity correlator,
\begin{equation}
\label{eq:noiseless-scaling}
    D^{(m)}_{ij}:=F(\rho^{(m)}, X_j X_i\rho^{(m)} X_i X_j).
\end{equation}
It is lower bounded by the noiseless correlator $D^{(m)}_{ij}\geq\tilde{D}^{(m)}_{ij}:=|\langle\psi^{(m)}|X_i X_j|\psi^{(m)}\rangle|$, with equality holding if $|\psi^{(m)}\rangle$ is sign-free in the $Z$ basis~\cite{weinstein2025efficient}. For sign-free cases, our algorithm also yields $D^{(m)}_{ij} = \tilde{D}^{(m)}_{ij}$ using a trivial optimal circuit (see SM for details). This includes the ground state ($m=0$) and the first excited state ($m=1$).

\begin{figure}
    \centering
    \includegraphics[width=\linewidth]{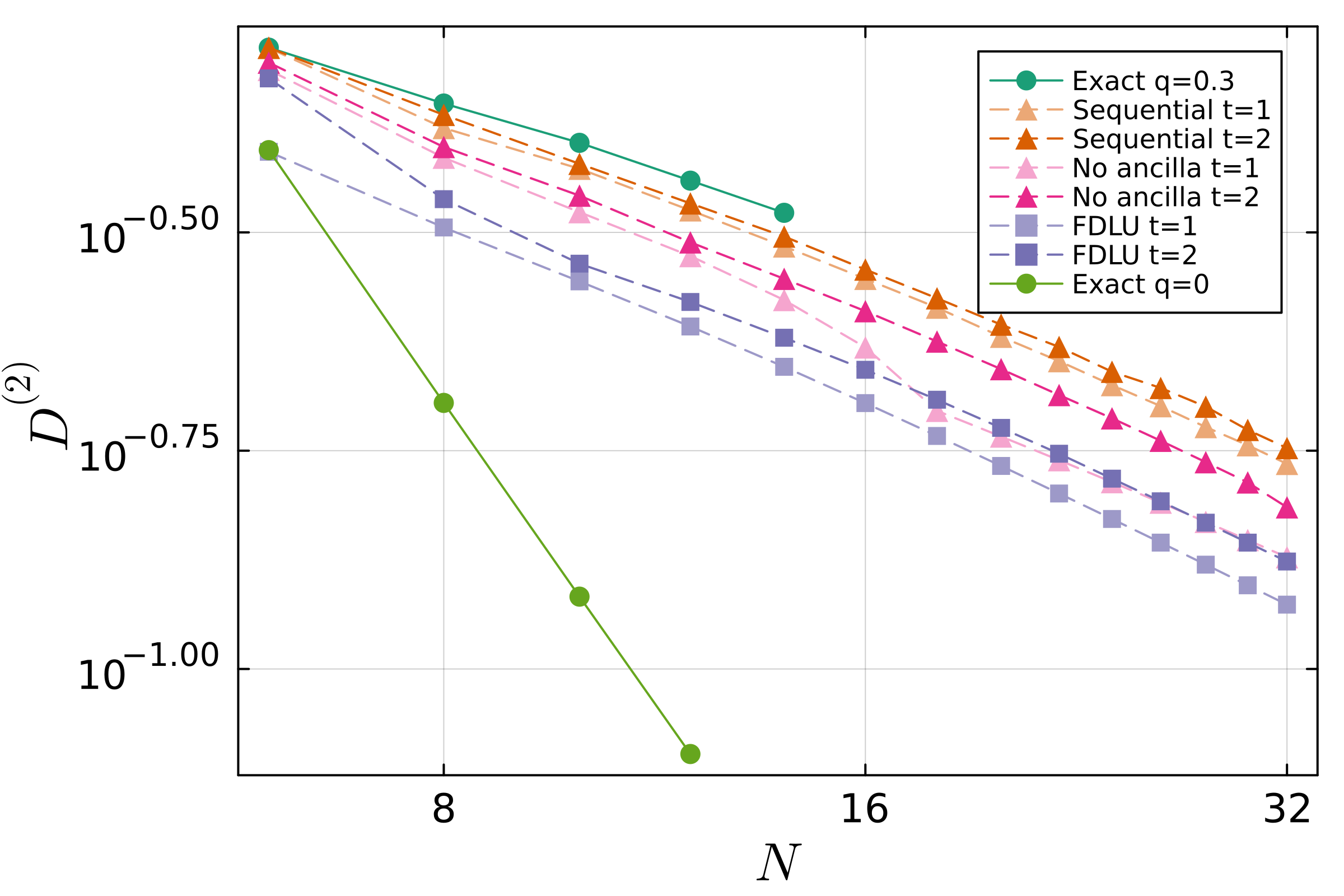}
    \caption{The fidelity correlator $D^{(2)}_{ij}$ of the second excited state of the critical Ising model under dephasing. We fix $|i-j|=N/2$ and vary the system size $N$. The $q=0$ solid line represents the noiseless correlator $\tilde{D}^{(2)}_{ij}$ which decays as $N^{-9/4}$. The $q=0.3$ solid line represents the exact fidelity correlator of the dephased state up to $N=14$. The dashed lines represent the lower bounds for the $q=0.3$ case obtained through different circuit structures, including sequential circuits, sequential circuits without ancilla, and FDLU circuits (with ancilla). For FDLU circuits, a depth-$1$ circuit includes one layer of even unitaries and one layer of odd unitaries. The lower bounds of $D^{(2)}$ consistently decay as $N^{-0.6}$, where the exponent is consistent with the extrapolation from the exact values with $N\leq 14$. }
    \label{fig:fidelity-correlator}
\end{figure}

To demonstrate the non-trivial utility of our method, we consider the second excited state ($m=2$) at criticality ($h=1$), which is long-range correlated and not sign-free. For the antipodal points ($|i-j| = N/2$), the noiseless correlator scales as $\tilde{D}^{(2)}_{ij} \propto N^{-9/4}$, where the exponent is the scaling dimension of a specific CFT operator  (see SM for details). However, under dephasing noise with strength $q=0.3$, our certified lower bound indicates that the fidelity correlator $D^{(2)}_{ij}$ follows a much slower power-law decay, as shown in Fig.~\ref{fig:fidelity-correlator}. Furthermore, while the sequential circuit with $t=2$ gives the best lower bound, different circuit structures yield similar power laws $D^{(2)}_{ij}\propto N^{-0.6}$. This power law is also consistent with the finite-size extrapolation of the exact fidelity.  Our numerical result thus highlights the change of the critical exponent under dephasing, indicating a different universality class for the mixed state. We emphasize the polynomial-time efficiency of our approach: the $N=32$ bound with depth $t=1$ is obtained in only 10 minutes on a laptop.

\textit{Codeword distinguishability --} Our method is also useful to study the robustness of logical information in approximate QEC codes. Given a code subspace $\mathcal{H}_c$, a noise channel $\mathcal{N}$ can be corrected only if the codeword states are still distinguishable. Consider orthogonal codewords $\rho, \sigma \in \mathcal{H}_c$ with $\tr\rho\sigma = 0$, we can compute $F(\mathcal{N}(\rho),\mathcal{N}(\sigma))$ as an indicator of error correctability -- decodable errors necessarily require $F(\mathcal{N}(\rho),\mathcal{N}(\sigma)) = 0$. In the following, we study this quantity for the recently proposed CFT code \cite{yi2024complexity,sang2024approximate,zhang2025probingmixedstatephasesquantum}, where the code subspace is the low-energy subspace of the critical Ising model. For $\mathcal{N}$ being uniform $X$ or $Z$ dephasing with strength $0<q<1$, it has been shown that the former is uncorrectable, while the latter is correctable in the thermodynamic limit. To further corroborate this result, we consider two codewords $\rho = |\psi^{(0)}\rangle \langle \psi^{(0)}|$ and $\sigma = |\psi^{(2)}\rangle\langle \psi^{(2)}|$ under the dephasing noise. For $X$ dephasing, the fidelity is lower bounded by $O(1)$ as $N$ increases, consistent with non-decodable noise \cite{sang2024approximate}. For $Z$ dephasing which is decodable, we show that the fidelity follows a universal scaling as
\begin{equation}
\label{eq:scaling_main}
    F(\mathcal{N}(\rho),\mathcal{N}(\sigma)) \propto N^{x},~~ x=-1/2
\end{equation}
where the exponent is determined by the scaling dimension of the energy operator $\Delta_{\varepsilon} = 1$ through $ x =(1-2\Delta_{\varepsilon})/2$. The argument is based on a scaling form of the fidelity similar to Ref.~\cite{sang2024approximate} (see details in SM). 

\begin{figure}
    \centering
    \includegraphics[width=\linewidth]{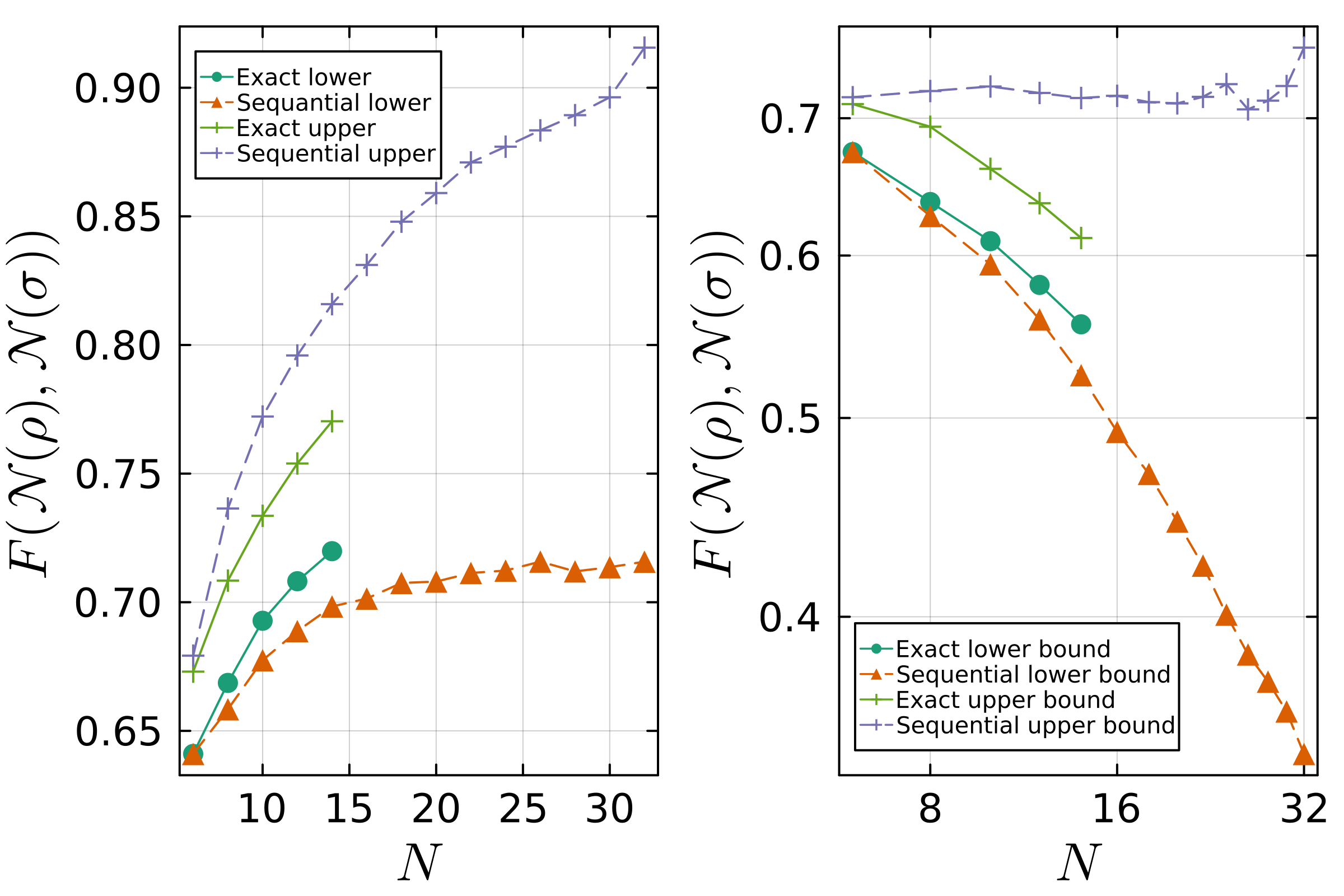}
    \caption{Fidelity $F(\mathcal{N}(\rho), \mathcal{N}(\sigma))$ of the ground state and second excited state of the critical Ising model under $X$ (left) or $Z$ (right) dephasing noise with strength $q=0.3$. The lower bound on the left suggests that the noise is undecodable. The lower bound on the right shows power law $N^{-1/2}$ behavior, where the exponent agrees with the scaling analysis of the true fidelity, indicating $N$-independent relative error. We also plot the upper bound through Eq.~\eqref{eq:vandegraaf} and find that they provide stronger bounds than the superfidelity.}
    \label{fig:cft-code}
\end{figure}

As shown in Fig.~\ref{fig:cft-code}, we compute bounds of $F(\mathcal{N}(\rho),\mathcal{N}(\sigma))$ obtained through the $t=2$ sequential circuit up to $N=32$. For $X$ dephasing, both the lower bound and upper bound increase with $N$, consistent with undecodable error. For $Z$ dephasing, we observe that the lower bound decays with $N^{-1/2}$, correctly capturing the universal scaling behavior Eq.~\eqref{eq:scaling_main}. It is remarkable that although the variational bound remains quantitatively distinct from the exact fidelity, it faithfully captures the underlying universal physics. For $N = 14$, where exact results are available, the relative error of the lower bound is approximately $5.5\%$. The persistence of the $N^{-1/2}$ scaling at larger $N$ suggests that this relative error remains stable, which highlights the scalability of our approach. Finally, we also compute the certified upper bound through Eq.~\eqref{eq:vandegraaf} and find that they provide better bounds than the superfidelity.

\textit{Conclusion--}
We have developed a polynomial-time algorithm for computing certified lower and upper bounds on the fidelity between many-body mixed states represented as MPDOs. By optimizing the fidelity over sequential circuits with ancilla, we overcome the exponential scaling of exact diagonalization, reaching system sizes up to $N=32$. We demonstrated its power through applications to fidelity correlators in decohered critical states and codeword distinguishability in the CFT code. For the former application, we discover the change of critical exponent for the fidelity correlator under dephasing, indicating a different universality class for the decohered state. For the latter application, we find that the scaling of the fidelity lower bound agrees with field-theoretic arguments, indicating system-size stability of the estimation error.  

This work opens up several interesting research directions. First,  we observed that even a $t=1$ sequential circuit provides remarkably accurate lower bounds. Understanding the scaling of the variational error with depth $t$ could reveal the complexity of the optimal purification. It would also be interesting to understand why the lower bound gives an accurate critical exponent, which is perhaps due to an interplay between mixed-state renormalization group~\cite{sang2024mixed,sang2024approximate,zou2023channeling} and the sequential circuit ansatz. 
Second, the sequential circuit framework naturally extends to higher-dimensional tensor networks such as projected entangled pair operators (PEPOs), enabling the study of mixed-state phases in two-dimensional systems. Third, our variational approach can be adapted to other information-theoretic quantities. Notably, Uhlmann's theorem has been recently generalized to $\alpha$-Renyi relative entropies~\cite{Mazzola_2025}. Our method could provide efficiently computable variational bounds on these entropies, which can be subsequently tightened using inter-entropy inequalities. Finally, while presented here as a classical method, our method can be naturally implemented on quantum hardware through variational quantum algorithms \cite{Cerezo_2021_review}, enabling experimental certification of large-scale mixed states.

\textit{Acknowledgment--} Y.Z. thanks in particular Yuxuan Zhang for collaboration on a related topic. We thank Armando Bellante, Ignacio Cirac, Jeongwan Haah, Shengqi Sang, Georgios Styliaris, Guifre Vidal, Chong Wang for helpful discussions. Y.L. is supported by the Alexander von Humboldt Foundation. This work was supported by the Perimeter Institute for Theoretical Physics (PI) and the Natural Sciences and Engineering Research Council of Canada (NSERC). Research at PI is supported in part by the Government of Canada through the Department of Innovation, Science and Economic Development Canada and by the Province of Ontario through the Ministry of Colleges and Universities.

\textit{Codebase availability --} The code used for this project is public on github at \href{https://github.com/yijian-physics/MPDO-fidelity}{this link}.

\bibliography{ref}

\newpage
\onecolumngrid
\appendix

\section{Fidelity Correlator for the Critical Ising Model}
\label{sec:appA}
In this section, we provide more details on the fidelity correlator of the critical Ising model in the analytically tractable cases.

\subsection{Noiseless correlator: CFT analysis}
The low-energy eigenstates of the critical spin chain are in one-to-one correspondence with the operators in the CFT \cite{Blote86,Cardy_1984}. For the Ising model, the ground state $(m=0)$, first excited state $(m=1)$ and the second excited state $(m=2)$ correspond to the three primary operators $\mathbb{I}, \sigma, \varepsilon$, respectively. In the noiseless limit, the fidelity correlator is the expectation value:
\begin{equation}
\label{eq:appnoiseless}
    \tilde{D}^{(m)}_{ij} = |\langle \psi^{(m)}| X_i X_j |\psi^{(m)}\rangle |.
\end{equation}
We will analyze the scaling behavior of this correlator using CFT. First, we need the expansion of lattice operators $X$ in terms of the CFT operators \cite{CARDY1986186,Zou_2020}
\begin{equation}
\label{eq:appCFTexpansion}
    X = a_{\sigma} \sigma + a_{\partial^2_{\tau} \sigma} \partial^2_{\tau} \sigma + a_{\partial^2_{x} \sigma} \partial^2_{x} \sigma + \cdots
\end{equation}
where $\cdots$ represent operators with higher scaling dimensions. 
The coefficients $a_{\sigma}, a_{\partial^2_{\tau} \sigma}, a_{\partial^2_{x} \sigma} $ are nonzero, and the precise values are computed in Ref.~\cite{Zou_2020}.

The expansion allows us to expand the correlator as a CFT four-point correlation function on the complex plane. First, we substitute Eq.~\eqref{eq:appCFTexpansion} into Eq.~\eqref{eq:appnoiseless} and map to the complex plane,
\begin{equation}
    \tilde{D}^{(m)}_{ij} = a^2_{\sigma} \left(\frac{2\pi}{N}\right)^{2\Delta_{\sigma}}\langle \psi^{(m)}| \sigma(1) \sigma(e^{2\pi \text{i} (i-j)/N})|\psi^{(m)}\rangle + O(N^{-2\Delta_\sigma-2}),
\end{equation}
where $\Delta_{\sigma} = 1/8$ and the second term comes from the $\partial^{2}_{\tau}\sigma$ and $\partial^{2}_{x}\sigma$ terms in the expansion. Second, we restrict the state to be primary states. Using the state-operator correspondence we obtain 
\begin{equation}
    \langle\psi^{(m)}| \sigma(1) \sigma(e^{2\pi \text{i} (i-j)/N})|\psi^{(m)}\rangle =   \langle \psi^{(m)}(0) \sigma(1) \sigma(z,\bar{z}) \psi^{(m)}(\infty)\rangle
\end{equation}
where $\psi^{(m)}(\infty) := \lim_{T\rightarrow\infty} T^{2\Delta_m}\psi^{(m)}(T)$ and $z = e^{2\pi \text{i} (i-j)/N}$ is the complex coordinate. For the special case of $i-j=N/2$ we have $z = \bar{z} = -1$. Lastly, we recall the form of four-point correlation functions for the Ising CFT \cite{BELAVIN1984333,Mattis_1987},
\begin{equation}
    \begin{aligned}
        \langle \mathbb{I}(0) \sigma(1) \sigma(z,\bar{z}) \mathbb{I}(\infty)\rangle &= \frac{1}{|1-z|^{2\Delta_{\sigma}}} \\
    \langle \sigma(0) \sigma(1) \sigma(z,\bar{z}) \sigma(\infty)\rangle &= \frac{1}{2|z|^{1/4} |1-z|^{1/4}}(|1+\sqrt{1-z}|+|1-\sqrt{1-z}|)\\
    \langle \varepsilon(0) \sigma(1) \sigma(z,\bar{z}) \varepsilon(\infty)\rangle &= \frac{1}{4 |1-z|^{1/4} |z|}|1+z|^2.
    \end{aligned}
\end{equation}
We note that the last correlator \textit{vanishes} at $z=\bar{z} = -1$. This induces a distinct scaling behavior for the correlator $\tilde{D}^{(2)}_{ij}$ at $|i-j|=N/2$. The correlator scales as $O(N^{-9/4})$, coming from the second term in Eq.~\eqref{eq:appCFTexpansion}. For $m=0,1$ we have $\tilde{D}^{(m)}_{ij}= O(N^{-1/4})$. From the wavefunction point of view, the second excited state has a nontrivial sign structure such that the leading order piece precisely cancels as we insert two $X$ operators at antipodal points. 
\subsection{Fidelity correlators for sign-free states under dephasing}
Here we prove that for \textit{any} $Z$-dephasing strength $0\leq q \leq 1$, $D^{(m)}_{ij} = \tilde{D}^{(m)}_{ij}$ if $|\psi^{(m)}\rangle$ is a sign-free state in $Z$ basis.  We note that $|\psi^{(m)}\rangle$ for $m=0,1$ satisfies the sign-free condition, i.e., $|\psi^{(m)}\rangle$ can be expanded using the eigenbasis $\{|z\rangle\}$ of $Z$ such that
\begin{equation}
    |\psi\rangle=\sum_z c_z |z\rangle,\quad c_z\geq 0.
\end{equation}
We first focus on the fully dephasing case $q=1$. The $Z$-dephased density matrix is then diagonal 
\begin{eqnarray}
    \rho=\prod_i\mathcal{N}_{i}(|\psi\rangle\langle\psi|)=\sum_z c_z^2 |z\rangle\langle z|,
\end{eqnarray}
where the off-diagonal terms vanish because for any term $|z\rangle\langle z'|$ with $|z\rangle\neq |z'\rangle$, there exists a position $k$ such that $\mathcal{N}_{k}(|z\rangle\langle z'|)=\frac{1}{2}(|z\rangle\langle z'|+Z_k |z\rangle\langle z'| Z_k)=0$. Therefore, 
\begin{equation}
\begin{aligned}
    \sigma:&=X_i X_j \rho X_i X_j \\
    &= \sum_z c_z^2 X_i X_j|z\rangle\langle z|X_i X_j=\sum_z c_z^2 |\tilde{z}\rangle\langle \tilde{z}|,
\end{aligned}
\end{equation}
where $|\tilde{z}\rangle:=X_i X_j |z\rangle$ and $\{|\tilde{z}\rangle\}$ is still the eigenbasis of $Z$'s. Since $c_z\geq 0$, one can take the square root in a trivial way 
\begin{eqnarray}
\begin{aligned}
    \sqrt{\rho}&=\sum_z c_z |z\rangle\langle z|,\\
    \sqrt{\sigma}&=\sum_z c_z |\tilde{z}\rangle\langle\tilde{z}|.
\end{aligned}
\end{eqnarray}
Furthermore, for the $q=1$ case, the states $\rho,\sigma$ are are both diagonal in the $Z$ basis and commute, thus
\begin{equation}
    D_{ij}^{(m)} :=F(\rho,\sigma):=\|\sqrt{\rho}\sqrt{\sigma}\|_1= \tr \sqrt{\rho}\sqrt{\sigma} =\sum_z c_z c_{X_i X_j z},
\end{equation}
This establishes that 
\begin{equation}
    D_{ij}^{(m)} = |\langle \psi^{(m)}|X_i X_j|\psi^{(m)}\rangle|
\end{equation}
if $|\psi^{(m)}\rangle$ is sign-free in the $Z$ basis for the fully dephasing case $q=1$. Thus the fidelity correlator of $q=1$ exactly equals to that of $q=0$. For other values of dephasing strength $q$, we can easily show that the fidelity correlator $F(\rho,\sigma)$ also equals the values at $q=0$ and $q=1$, because it is a monotonically non-decreasing function of $q$. To see this, we note that
\begin{equation}
    \sigma = X_i X_j \left(\prod_k\mathcal{N}_{k}(|\psi\rangle\langle\psi|) \right)X_i X_j =  \left(\prod_k\mathcal{N}_{k}\right)(X_i X_j|\psi\rangle\langle\psi| X_i X_j) 
\end{equation}
because the channel $\mathcal{N}_{k}$ is \textit{weakly symmetric} under any bit flip $X_i$. Therefore, monotonicity of $F(\rho,\sigma)$ follows from monotonicity of the fidelity function (data processing inequality). As a result, we reach a simple conclusion for sign-free states that the $Z$ dephasing does not alter the fidelity correlator. 

\section{Toy examples of optimal fidelity}
In this section, we illustrate toy examples of optimal unitary $U$ that achieves the exact fidelity 
\begin{equation}
\label{eq:optimal_U_app}
        U=\text{argmax}_{U\in\mathrm{U}(\mathcal{H}_p)} |\dlangle \psi_\rho|(\bo\otimes U)|\psi_\sigma\drangle|,
\end{equation}
and can be written as a finite-$t$ sequential circuit with ancilla.
\subsection{Fidelity between MPS and MPDO}
\begin{figure}
    \centering
    \includegraphics[width=0.8\linewidth]{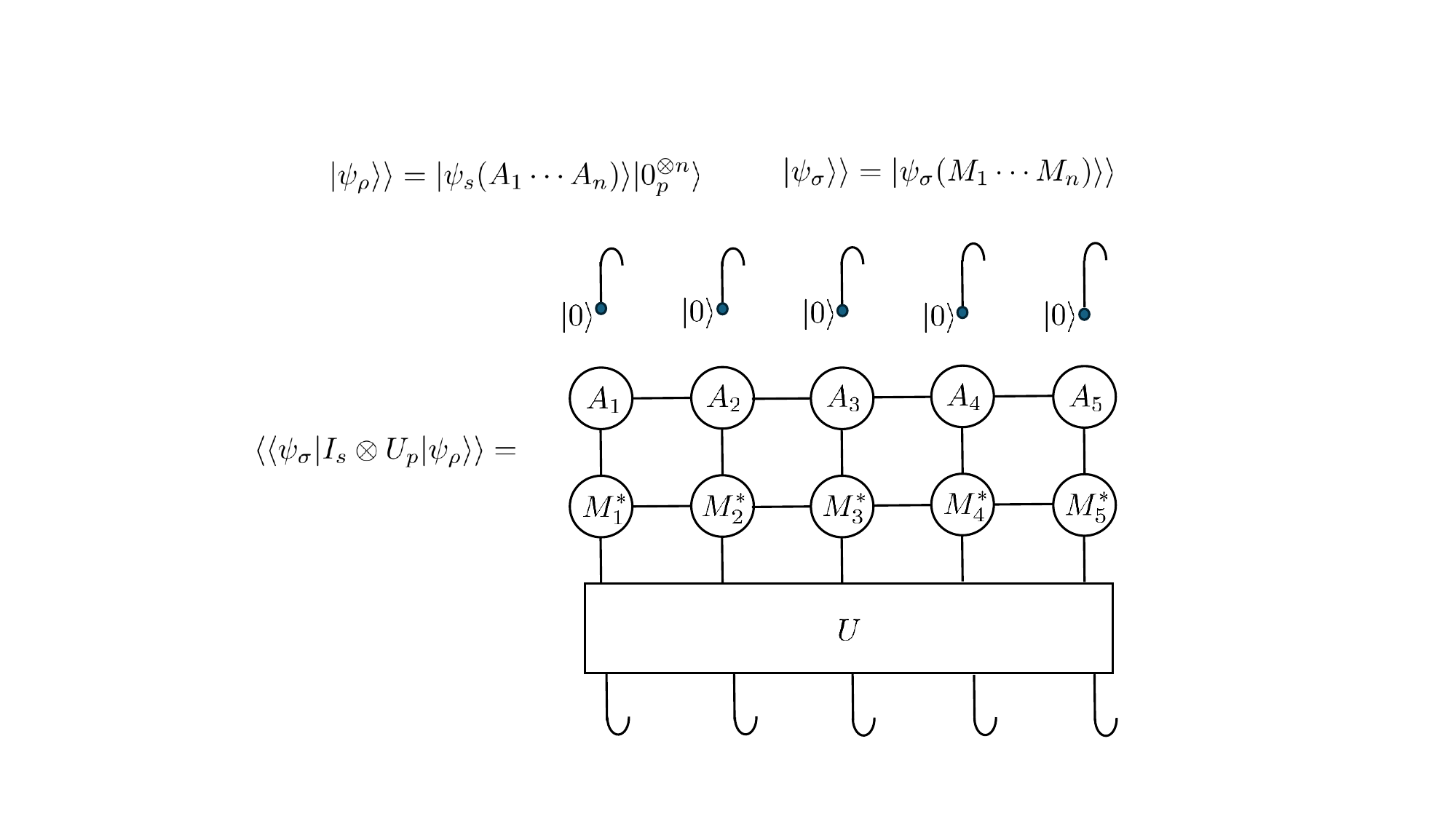}
    \caption{Optimal unitary of the fidelity between an MPS and a locally purified MPDO.}
    \label{fig:pure}
\end{figure}

We first consider $\rho = |\psi_s\rangle\langle \psi_s|$, where $|\psi_s\rangle = |\psi_s(A_1,\cdots A_n)\rangle$ is an MPS. The purification is given by $|\psi_\rho\drangle = |\psi_s\rangle|0^{\otimes n}_p \rangle$. The other state $\sigma$ is given as a locally purified MPDO $|\psi_{\sigma}\drangle$, with tensors $M_1,M_2,\cdots M_n$. The overlap Eq.~\eqref{eq:optimal_U_app} can be represented as the tensor network in Fig.~\ref{fig:pure}. Then the optimal $U$ is such that
\begin{equation}
    U|0^{\otimes n}\rangle  \propto  \langle \psi_s|\psi_{\sigma}\drangle
\end{equation}
The unnormalized state on the RHS is an MPS with finite bond dimension, thus can be prepared using a finite-depth sequential circuit, which is exactly the optimal $U$. The optimal fidelity is then
\begin{equation}
    F(\rho,\sigma) = ||\langle \psi_s|\psi_{\sigma}\drangle ||_2 = \sqrt{\langle\psi_s|\sigma|\psi_s\rangle}
\end{equation}
as expected.

\subsection{Logical operators of repetition code}
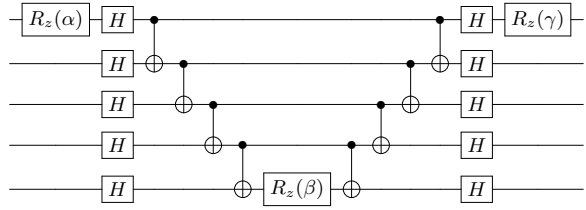
\begin{figure}[h]
\centering
\scalebox{0.85}{
\Qcircuit @C=0.6em @R=0.7em {
& \gate{R_z(\alpha)} & \gate{H} & \ctrl{1} & \qw & \qw & \qw & \qw & \qw & \qw & \qw & \ctrl{1} & \gate{H} & \gate{R_z(\gamma)} & \qw \\
& \qw & \gate{H} & \targ & \ctrl{1} & \qw & \qw & \qw & \qw & \qw & \ctrl{1} & \targ & \gate{H} & \qw & \qw \\
& \qw & \gate{H} & \qw & \targ & \ctrl{1} & \qw & \qw & \qw & \ctrl{1} & \targ & \qw & \gate{H} & \qw & \qw \\
& \qw & \gate{H} & \qw & \qw & \targ & \ctrl{1} & \qw & \ctrl{1} & \targ & \qw & \qw & \gate{H} & \qw & \qw \\
& \qw & \gate{H} & \qw & \qw & \qw & \targ & \gate{R_z(\beta)} & \targ & \qw & \qw & \qw & \gate{H} & \qw & \qw
}
}
\caption{Sequential circuit realization of logical rotation $\bar{U} = \bar{R}_z(\alpha) \bar{R}_x(\beta)\bar{R}_z(\gamma)$ on the repetition code subspace.}
\label{fig:GHZ}
\end{figure}

Consider two mixed states $\rho,\sigma$ in the logical subspace $\mathcal{H}_L$ of the repetition code spanned by $|\bar{0}\rangle :=|0^{\otimes N}\rangle$ and $|\bar{1}\rangle :=|1^{\otimes N}\rangle$. Let $\rho = \sum_{i,j=1}^2 \bar{\rho}_{ij} |\bar{i}\rangle \langle \bar{j}|$ and $\sigma = \sum_{i,j=1}^2 \bar{\sigma}_{ij} |\bar{i}\rangle \langle \bar{j}|$, then the fidelity is $F(\rho,\sigma) = F(\bar{\rho},\bar{\sigma})$. Given any purification $|\psi_{\rho}\drangle, |\psi_{\sigma}\drangle \in \mathcal{H}_L \otimes \mathcal{H}_L $,  the optimal unitary $U$ must perform a logical operation. To see this, we can expand
\begin{equation}
    |\psi_{\rho}\drangle = \sum_{i,j=1}^2 A_{ij} |\bar{i}\rangle |\bar{j}\rangle, ~ |\psi_{\sigma}\drangle = \sum_{i,j=1}^2 B_{ij} |\bar{i}\rangle |\bar{j}\rangle
\end{equation}
where $\bar{\rho} = AA^{\dagger}, \bar{\sigma} = BB^{\dagger}$. Then $F(\rho,\sigma) = ||A^{\dagger}B||_1$.  Consider the unitary $\bar{U} \in SU(2)$ such that $|\tr(\bar{U}BA^{\dagger})| = ||BA^{\dagger}||_1$, then the unitary $U$ such that 
\begin{equation}
    U|\bar{j}\rangle = \sum_{i} \bar{U}_{ij} |\bar{i}\rangle
\end{equation}
achieves the optimal unitary. This logical operation can be realized by a $t=1$ sequential unitary without ancillas. More specifically, consider the Euler angle decomposition
\begin{equation}
    \bar{U} = \bar{R}_z(\alpha) \bar{R}_x(\beta)\bar{R}_z(\gamma)
\end{equation}
where $\bar{R}_z(\alpha)  = R_{z,1}(\alpha) = e^{-i\alpha Z_1/2}$ and $\bar{R}_x(\beta) = e^{-i\beta X_1X_2\cdots X_N/2}$. The rotation $\bar{R}_x$ can be realized by Hadmard gates and two CNOT ladders, as shown in Fig.~\ref{fig:GHZ}. Finally, one can absorb the one-qubit gates into the CNOT ladders to form a $t=1$ sequential circuit, realizing the logical operator $U$ that implements $\bar{U}$ in the code subspace.

\subsection{Sign-free states under dephasing}
In this section, we consider two sign-free states 
\begin{equation}
    |\psi_1\rangle = \sum_{z} a_z|z\rangle, ~|\psi_2\rangle = \sum_{z} b_z|z\rangle
\end{equation}
where $a_z,b_z\geq 0$ and $\{|z\rangle\}$ is the computational basis. Now we consider the noise channel $\mathcal{N}=\otimes_i \mathcal{N}_{i,(q)}$, where $\mathcal{N}_{i,(q)}$ is the $Z$-dephasing channel acting on site $i$ with strength $q$. The two mixed states are $\rho = \mathcal{N}(|\psi_1\rangle\langle \psi_1|)$ and $\sigma = \mathcal{N}(|\psi_2\rangle\langle \psi_2|)$.

As we have shown in Sec.~\ref{sec:appA}, montonocity of fidelity and the sign-free condition ensures that
\begin{equation}
    F(\rho,\sigma) = |\langle \psi_1|\psi_2\rangle|
\end{equation}
independent of the noise rate $q$. This also indicates that the optimal fidelity can be found with a trivial circuit. The locally purified MPDOs can be easily constructed by
\begin{equation}
    |\psi_{\rho}\drangle = (\otimes_{i} U_{s_i,p_i})|\psi_1\rangle_s |0^{\otimes n}\rangle_p, ~~ |\psi_{\sigma}\drangle = (\otimes_{i} \tilde{U}_{s_i,p_i})|\psi_2\rangle_s |0^{\otimes n}\rangle_p
\end{equation}
where $U_{s_i,p_i}$ and $\tilde{U}_{s_i,p_i}$ are two Steinspring dilations of the noise channel $\mathcal{N}_i^{(q)}$ on site $i$. The two dilations are related by $U_{s_i,p_i} =V_{p_i} U_{s_i,p_i}$, where $V_{p_i}$ is a unitary acting on the purification spin $i$. It follows that
\begin{equation}
    \dlangle \psi_{\rho}| \otimes_i V_{p_i}|\psi_{\sigma}\drangle = |\langle \psi_1|\psi_2\rangle| = F(\rho,\sigma). 
\end{equation}
As a result, the optimal unitary is given by a product unitary $U=\otimes_i V_{p_i}$.

\subsection{Realization of matrix product unitaries}
In this section, we consider the scenario where the two mixed states are the same $\rho=\sigma$ but with different purifications $|\psi_{\rho}\drangle$ and $|\psi_{\sigma}\drangle$. Let $|\psi_{\sigma}\drangle$ be a finite bond-dimension two-sided MPS, and $|\psi_{\rho}\drangle$ is related to it by a matrix product unitary $U$ acting on the purifying space.
\begin{equation}
    |\psi_{\rho}\drangle  = (I\otimes U)|\psi_{\sigma}\drangle
\end{equation}
If $U$ has finite bond dimension, then it is automatically guaranteed that $|\psi_{\rho}\drangle$ has a finite bond dimension. In this way, we have two MPDO representations of the same state, and the optimal unitary that gives $F(\rho,\sigma) = 1$ is given exactly by the MPU. We will check whether the optimal unitary can be represented by the sequential circuit ansatz.

All uniform (translation-invariant) MPUs can be written as a composition of FDLUs and translations by a constant amount \cite{cirac2017matrix}. Note that both FDLUs and translations can be represented as a finite-$t$ sequential circuit; our ansatz is sufficient for this case. Note that this class of circuit belongs to quantum cellular automata (QCA), which is defined as locality-preserving unitary maps in the operator space.

\begin{figure}
    \centering
    \scalebox{0.8}{ 
            \Qcircuit @C=0.8em @R=0.5em {
                & \ctrl{1}  & \qw  & \qw & \qw  & \qw& \qw  & \ctrl{1}  & \qw \\
                \lstick{\ket{0}} 
                & \gate{X} & \ctrl{1} & \qw  & \qw & \qw & \ctrl{1}  & \gate{X} & \qw & \rstick{\ket{0}} \\
                & \qw  & \ctrl{1}  & \qw & \qw & \qw & \ctrl{1} & \qw & \qw\\
                \lstick{\ket{0}} 
                & \qw & \gate{X} & \ctrl{1} & \qw & \ctrl{1} & \gate{X}  & \qw & \qw & \rstick{\ket{0}} \\
                & \qw & \qw  & \ctrl{1} & \qw  & \ctrl{1} & \qw  & \qw & \qw\\
                \lstick{\ket{0}} 
                & \qw & \qw & \gate{X} & \ctrl{1} & \gate{X}  & \qw  & \qw & \qw & \rstick{\ket{0}} \\
                & \qw & \qw & \qw & \gate{U}& \qw & \qw & \qw& \qw
            }
            }
    \caption{Decomposition of multi-controlled-$U$ gate $U_{\text{MCU}}$ on $N=4$ qubits by $t=1$ sequential circuit with ancilla. }
    \label{fig:mcz}
\end{figure}
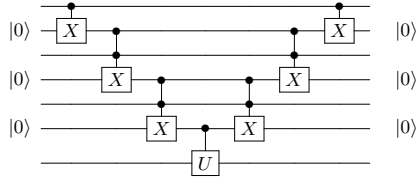

\begin{figure}
    \centering
    \includegraphics[width=0.3\linewidth]{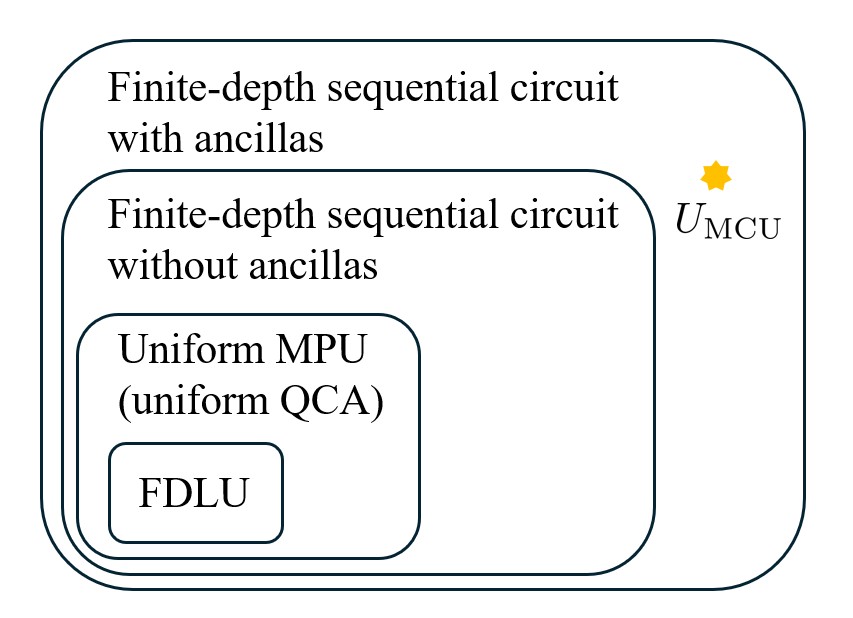}
    \caption{Expressiveness of different ansätze for unitary. }
    \label{fig:hierachy}
\end{figure}

Next, we examine non-uniform MPUs. Up to now, there is no complete classification on non-uniform MPUs. However, it is known that certain non-uniform MPUs go beyond QCA \cite{styliaris2025matrix}, such as the multi-controlled-$U$ operator
\begin{equation}
\label{eq:MCU_app}
    U_{\text{MCU}}=\bo_2^{\otimes N}+(|1\rangle\langle 1|)^{\otimes (N-1)} \otimes (U-\bo_2),
\end{equation}
This unitary implements a $U$ transformation on the last spin if all other spins are in $|1\rangle$ and implements the identity operation otherwise. We can use a $t=1$ sequential circuit with ancillas to realize it, as shown in Fig.~\ref{fig:mcz}. Here the ancilla are used for recording whether the control spins are in all $|1\rangle$ states. We summarize the expressiveness of different ansätze for unitary in Fig.~\ref{fig:hierachy}. 

\section{Codeword distinguishability for CFT code}
In this section, we consider the CFT code that involves the low-energy states $|\psi^{(m)}\rangle$ of the critical Ising model $H = -\sum_{i=1}^{N} X_i X_{i+1} - \sum_{i=1}^{N} Z_i$. Let the dephased eigenstates be
\begin{equation}
    \rho^{(m)}_{q} := \mathcal{N}(|\psi^{(m)}\rangle\langle \psi^{(m)}|),
\end{equation}
where $\mathcal{N}$ is a uniform $Z$-dephasing with rate $0<q<1$ and $|\psi^{(m)}\rangle$ is the $m$-th lowest eigenstate of the critical Ising model.

We will show that the fidelity of dephased eigenstates of the Ising CFT satisfies that
\begin{equation}
\label{eq:scaling_res}
    F(\rho^{(0)}_q, \rho^{(2)}_q) = O(N^{-1/2}),
\end{equation}
with a fixed $q$ in the thermodynamic limit.
Our argument is similar to the scaling form of coherent information in Ref.~\cite{sang2024approximate}. Specifically, we assume that the fidelity is only a function of $qN^{\nu}$, where $\nu$ is a critical exponent that controls the RG flow of the noise rate.
\begin{equation}
\label{eq:scaling}
    F(\rho^{(0)}_q, \rho^{(2)}_q) =  f(qN^{\nu})
\end{equation}
For the correctable error in the thermodynamic limit, we have $\nu<0$ and $f(0)=0$. Furthermore, we assume that the trace distance $T(\rho,\sigma):=\frac{1}{2}\|\rho-\sigma\|_1$ is also controlled by the same critical exponent,
\begin{equation}
\label{eq:scalingT}
    T(\rho^{(0)}_q, \rho^{(2)}_q) =  1-t(qN^{\nu})
\end{equation}
with $t(0) = 0$. The assumption that the critical exponents are the same for both fidelity and trace distance is justified in the RG picture: $qN^{\nu}$ can be seen as the effective noise rate under the RG flow, which controls all the information-theoretic distances that govern decodability. The exponent therefore coincides with the same exponent $\nu = 1-2\Delta_Z = -1$ for the coherent information in Ref.~\cite{sang2024approximate}. The scaling form implies that taking $N\rightarrow \infty$ with fixed noise rate $q$ is equivalent to fixing $N$ and taking $q\rightarrow 0$. We will then work with the regime where $q$ is small and extract the leading term in the scaling function $f(x)$ and $t(x)$. 

In the limit of $q\rightarrow 0$, we expand
\begin{equation}
\label{eq:expansion_app}
   \rho^{(m)}_q= \left(1-\frac{qN}{2}\right) |\psi^{(m)}\rangle\langle \psi^{(m)}| + \frac{q}{2}\sum_{i=1}^{N}  Z_i |\psi^{(m)}\rangle\langle \psi^{(m)}| Z_i + O(q^2).
\end{equation}
It is then straightforward to compute that
\begin{equation}
    \tr \rho^{(0)}_q \rho^{(2)}_q = q  \sum_{i=1}^N|\langle \psi^{(0)}| Z_i |\psi^{(2)}\rangle|^2 + O(q^2)
\end{equation}
Note that since the scaling dimension $\Delta_Z = \Delta_{\varepsilon} = 1$, $|\langle \psi^{(0)}| Z_i |\psi^{(2)}\rangle| \propto N^{-\Delta_Z} = N^{-1}$, we then have
\begin{equation}
    \tr \rho^{(0)}_q \rho^{(2)}_q = O( qN^{1-2\Delta_Z})= O(qN^{-1}).
\end{equation}
This gives us a lower bound of the fidelity, since
\begin{equation}
    F(\rho^{(0)}_q, \rho^{(2)}_q) \geq \sqrt{\tr \left(\rho^{(0)}_q \rho^{(2)}_q\right)} = O(q^{1/2}N^{-1/2})
\end{equation}
Next, we compute an upper bound in terms of trace distance,
\begin{equation}
\label{eq:fuchs_appcft}
    F(\rho^{(0)}_q, \rho^{(2)}_q) \leq \sqrt{1-T(\rho^{(0)}_q, \rho^{(2)}_q)^2},
\end{equation}
Perturbation theory suggests that the eigenvalues of $\rho^{(0)}_q-\rho^{(2)}_q$ is linear in $q$ to the first order. Therefore, the leading term in $t(x)$ is linear, that is $t(x) = O(x)$. Thus, the upper bound with Eq.~\eqref{eq:fuchs_appcft} gives
\begin{equation}
    F(\rho^{(0)}_q, \rho^{(2)}_q) \leq O(q^{1/2}N^{-1/2})
\end{equation}
The lower and upper bound match each other, giving
\begin{equation}
    F(\rho^{(0)}_q, \rho^{(2)}_q) = O(q^{1/2}N^{-1/2})
\end{equation}
The scaling form Eq.~\eqref{eq:scaling} then implies that the consistent way is $f(x) = O(\sqrt{x})$. Note that we have derived the scaling form in the limit of $q\rightarrow 0$. As we explained earlier, the RG picture allows us to use the same scaling form with $q=O(1)$ and $N\rightarrow\infty$. Thus we have proven Eq.~\eqref{eq:scaling_res} for $q=O(1)$.

\begin{figure}
    \centering
    \includegraphics[width=0.5\linewidth]{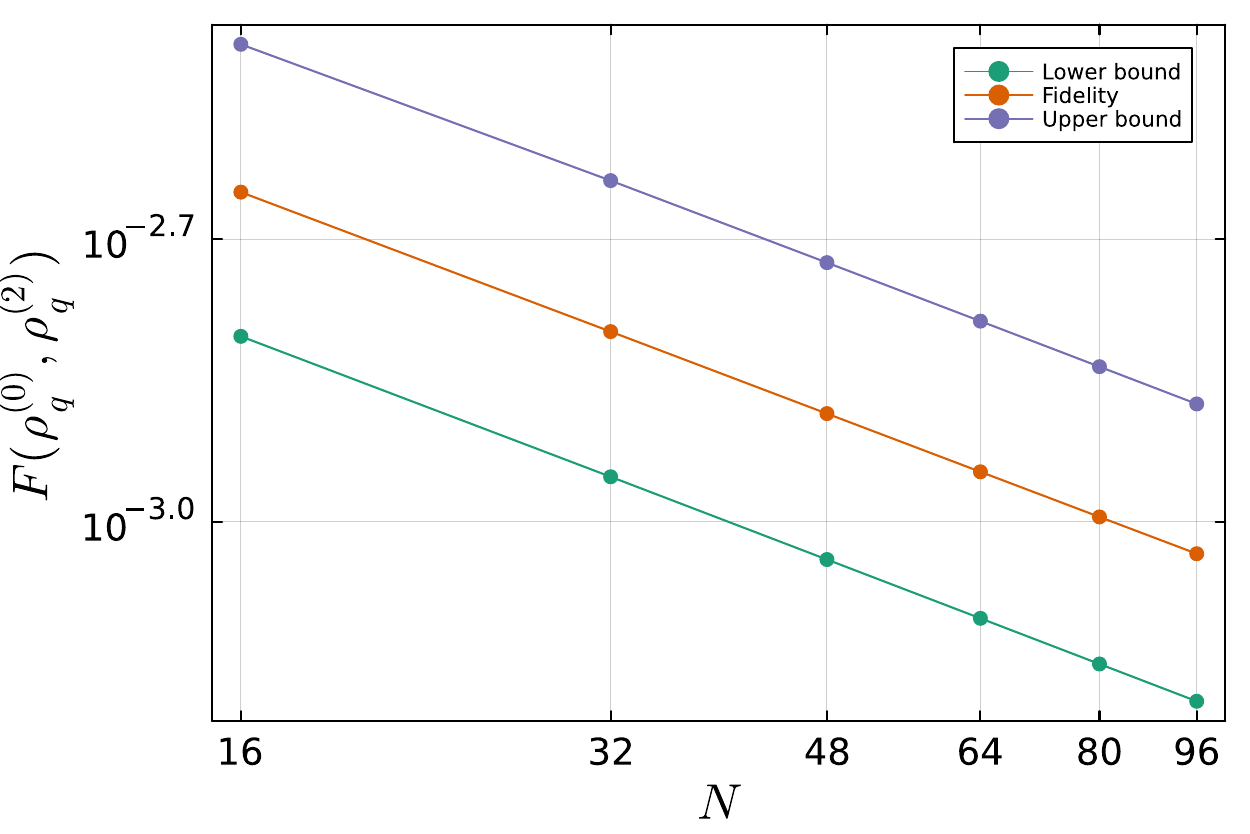}
    \caption{Fidelity $F(\rho^{(0)}_q, \rho^{(2)}_q)$, its lower bound $\sqrt{\tr \rho^{(0)}_q\rho^{(2)}_q}$ and upper bound $\sqrt{1-T(\rho^{(0)}_q, \rho^{(2)}_q)^2}$ in the perturbative regime $qN\ll 1$. We have fixed $q=10^{-5}$ and vary $N$ in the range $16\leq N \leq 96$. All three quantities decay as $N^{-1/2}$, in agreement with the scaling analysis.}
    \label{fig:perturbative}
\end{figure}

To further corroborate that the exponent $\nu = -1$ in the scaling form, we perform a numerical calculation of $F(\rho^{(0)}_q, \rho^{(2)}_q)$, its lower bound $\sqrt{\tr \rho^{(0)}_q\rho^{(2)}_q}$ and upper bound $\sqrt{1-T(\rho^{(0)}_q, \rho^{(2)}_q)^2}$ in the perturbative regime $qN\ll 1$. In this regime, $\rho_q^{(m)}$ in Eq.~\eqref{eq:expansion_app} has only rank $N+1$, which allows us to perform efficient numerical simulations. We have found that all three quantities decay as $N^{-1/2}$ in this regime, see Fig.~\ref{fig:perturbative}. This numerical results thus provide a self-consistency check of the proof above.

\end{document}